\documentstyle[aps,preprint,epsf]{revtex}

\begin{document}
\draft

\preprint{CERN-TH/99-155}
\title{QCD running coupling effects~ for the non-singlet
structure function at small $x$}

\author{B.I. Ermolaev\footnote
{Permanent address:
A.F. Ioffe Physico-Technical Institute, 194021 St.Petersburg, Russia}\\
CERN 1211, Geneve 23, Switzerland \\
 M. Greco\\
Dipartimento di Fisica and INFN, University of Rome III, Rome, Italy \\
and\\
S.I. Troyan\\
St.Petersburg Institute of Nuclear Physics, 188350 St.Petersburg-Gatchina,
Russia}

\maketitle

\begin{abstract}
A generalization of the leading-order DGLAP evolution at
small $x$ is performed for the non-singlet structure function
by resumming the leading-order DGLAP anomalous
dimension to all orders in the QCD coupling.
Explicit expressions are obtained for  the non-singlet
structure function of the deep
inelastic
scattering, taking into account both the double-logarithmic and the
single-logarithmic
contributions, including the running
$\alpha_s$ effects. It is shown that
when these contributions are included, the asymptotic small-$x$
behaviour is power-like, with an exponent of about 0.4.
\end{abstract}



\newpage

\section{Introduction}

Non-singlet structure functions, i.e. flavour-dependent contributions to the
deep inelastic structure functions, have been the object
of intensive theoretical
investigation. Besides being interesting in themselves, they are also
comparatively technically  simple for analysis, and can be
regarded
as a starting ground for a theoretical
description of DIS structure functions. The standard and the most
widespread
way for theoretical investigation of DIS structure functions is the DGLAP
evolution equations  \cite{dglap}. As is well known, in the
description of DIS, these equations
provide
also good agreement with experimental data \cite{a}
when they are used in the small- $x$ region,
although they are not expected to work very well
in this kinematical region. Actually, in the small-$x$ region, a
Regge-type behaviour for the DIS structure functions
cannot be obtained from DGLAP, which predicts
instead a dependence of the type
$f_{NS}  = e^{\sqrt{\ln(1/x) \ln(Q^2/\mu^2)}}$.
In perturbative QCD, such Regge-like behaviour was obtained
\cite{bfkl} a long
time ago for structure functions of unpolarized DIS and recently
\cite{emr,ber} for
structure functions of the polarized DIS.
However, all those results are obtained in the leading logarithmic
approximation,  which has the unsatisfactory
feature that $\alpha_s$ is regarded as a
constant.

In the present paper in addition to providing
formulae for the asymptotics of the non-singlet
structure function at very small $x$, we present also
an explicit expression for $f_{NS}$
for small but finite values of $x$, which accounts for both leading
(double-logarithmic) and subleading (single-logarithmic)
contributions to all orders in QCD coupling, including
running $\alpha_s$ effects.
As logarithms of $x$ and of $Q^2$ are
important at small $x$, we account for them both.
 In order to take  running $\alpha_s$ effects into account,
we use in the present work the approach
of \cite{kl}, with some important changes.
Earlier, the running $\alpha_s$ effects were incorporated into the
double-logarithmic equations for
the non-singlet structure function at small $x$
in \cite{emr,m}, but other single logarithmic contributions were not
taken into account in those works. In the present work we take them
into account, which makes our approach self-consistent.

In the DGLAP equations, the expressions for the anomalous dimensions
take into account a finite number of NLO contributions; in this approach,
the QCD coupling
depends on $Q^2$.
After the total resummation of the singular NLO
contributions to the anomalous dimensions with
the double logarithmic accuracy performed in \cite{emr,ber}, there appeared
works on the small-$x$ behaviour of the structure functions
combining, to a certain extent, the results of the DGLAP with
the results of \cite{emr,ber}.
In particular, the works \cite{b,jk,blum,kk}
treat the running QCD coupling at small $x$,
allowing $\alpha_s$ to depend on the incoming photon virtuality $Q^2$.
We arrive at a different conclusion.\\

The paper is organized as follows:
In Sect. 2 we derive the small-$x$ evolution equations for $f_{NS}$ and solve
them.
In Sect. 3 we obtain the asymptotic behaviour of $f_{NS}$ and discuss it.
Finally, Sect. 4 contains our conclusions.

\section{Small-$x$ evolution equation ~for non-singlet structure functions}

In the Born approximation, and suppressing unimportant factors, we define the
non-singlet structure function as

\begin{equation}
f_{NS}^{Born} = \delta(1- x).
\label{fborn}
\end{equation}

\begin{figure}
\begin{picture}(200,150)
\put(40,10){
\epsfbox{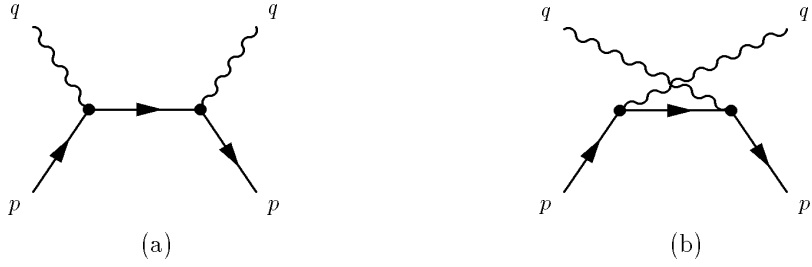}
}
\end{picture}
\caption{The Born graphs for the amplitude $M$.}
\label{Born}
\end{figure}

\begin{figure}
\begin{picture}(140,130)
\put(0,10){
\epsfbox{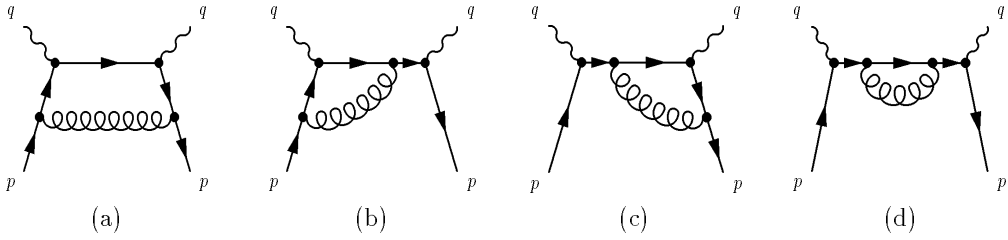}
}
\end{picture}
\caption{The graphs contributing to $f_{NS}$ in one-loop approximations.}
\label{first}
\end{figure}

For the radiative corrections to the Born approximation in the
kinematical region
of small $x$ and large $Q^2$, both logarithms of $Q^2$
and of $x$ are essential and should be taken into account. Therefore,
 a two-dimensional evolution equation is needed for $f_{NS}$, combining
the evolution with respect to  both $x$ and $Q^2$. It is simpler to obtain
such an equation for
the scattering amplitude $M(x, Q^2)$, with the $s$-channel
discontinuity (imaginary part) proportional to $f_{NS}$:

\begin{equation}
f_{NS} = \frac{1}{\pi}\Im_s M(x,Q^2),
\label{defM}
\end{equation}
where $s = (p + q)^2 = p^2 + 2pq (1 - x ) \approx 2pq$ is the
Mandelstam variable
corresponding to
the squared total energy of the process under discussion.
In the Born approximation, $M$ is represented by the
two graphs in Fig.\ref{Born}, although
only graph (a) has a non-zero imaginary part in $s$ and therefore only
that graph contributes to $f_{NS}$.
This is the reason why there is no difference
in the Born approximation between the non-singlet contributions
to the spin- dependent and to the spin-independent structure functions.
Graphs contributing to
$M$ at higher orders in  $\alpha_s$
are obtained from both graphs in Fig.\ref{Born}, by adding to them gluon
propagators and quark and gluon loops.
Starting from the order $\alpha_s^2$,
the graphs obtained from Fig.\ref{Born}b, with non-ladder gluon propagators
incorporated, contribute in different ways
to the non-singlet components of the spin-dependent and the
spin-independent structure functions (see \cite{ber}). In the present
paper we consider the non-singlet contribution to the spin-independet
structure functions $F_1$ and $F_2$.
In terms of the Regge theory we consider the positive signature contribution
to $f_{NS}$.
The first-loop correction to the Born graph (a) in Fig.\ref{Born} is represented by
graphs in Fig.\ref{first}. Having chosen the planar gauge

\begin{equation}
d_{\mu \nu} = g_{\mu \nu} + \frac{k_{\mu}n_{\nu} + n_{\mu}k_{\nu}}{nk}
\label{gauge}
\end{equation}
for the virtual gluon, we make the total contribution of all three
graphs in Fig.\ref{first} to be concentrated in the single graph (a). Iterating
this graph, we obtain all graphs (of the ladder type only) contributing
to $f_{NS}$.
In order to avoid the infrared (IR) singularities that appear in the radiative
corrections, we use Lipatov's prescription of compactifying the impact
parameter space (see e.g. \cite{l,kl,el}), in other words,
we introduce the
infrared
cut-off $\mu$
in the transverse space (with respect to the plane formed by $q$ and $p$) when
integrating over the momenta of virtual particles. With such a cut-off
acting as a
mass scale, one can neglect quark masses and still be free from IR
singularities.
Until now the procedure is not different from the one used
in the DGLAP, apart from the infrared regularization, which is a
purely technical
issue. The main difference comes  when
specifying the limits of integration over the transverse momenta of virtual
particles. In the DGLAP, all ladder transverse momenta $k_{i \perp}$ are
ordered as

\begin{equation}
k_{1 \perp}\leq k_{2 \perp}\leq ...\leq Q^2,
\label{order}
\end{equation}
with the numeration running from the bottom of the ladder to the top.\\

\begin{figure}
\begin{center}
\begin{picture}(140,200)
\put(0,10){
\epsfbox{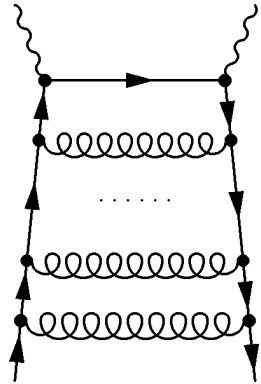}
}
\end{picture}
\end{center}
\caption{A typical graph for $f_{NS}$ in the planar gauge.}
\label{ladder}
\end{figure}

\begin{figure}
\begin{picture}(300,340)
\put(0,10){
\epsfbox{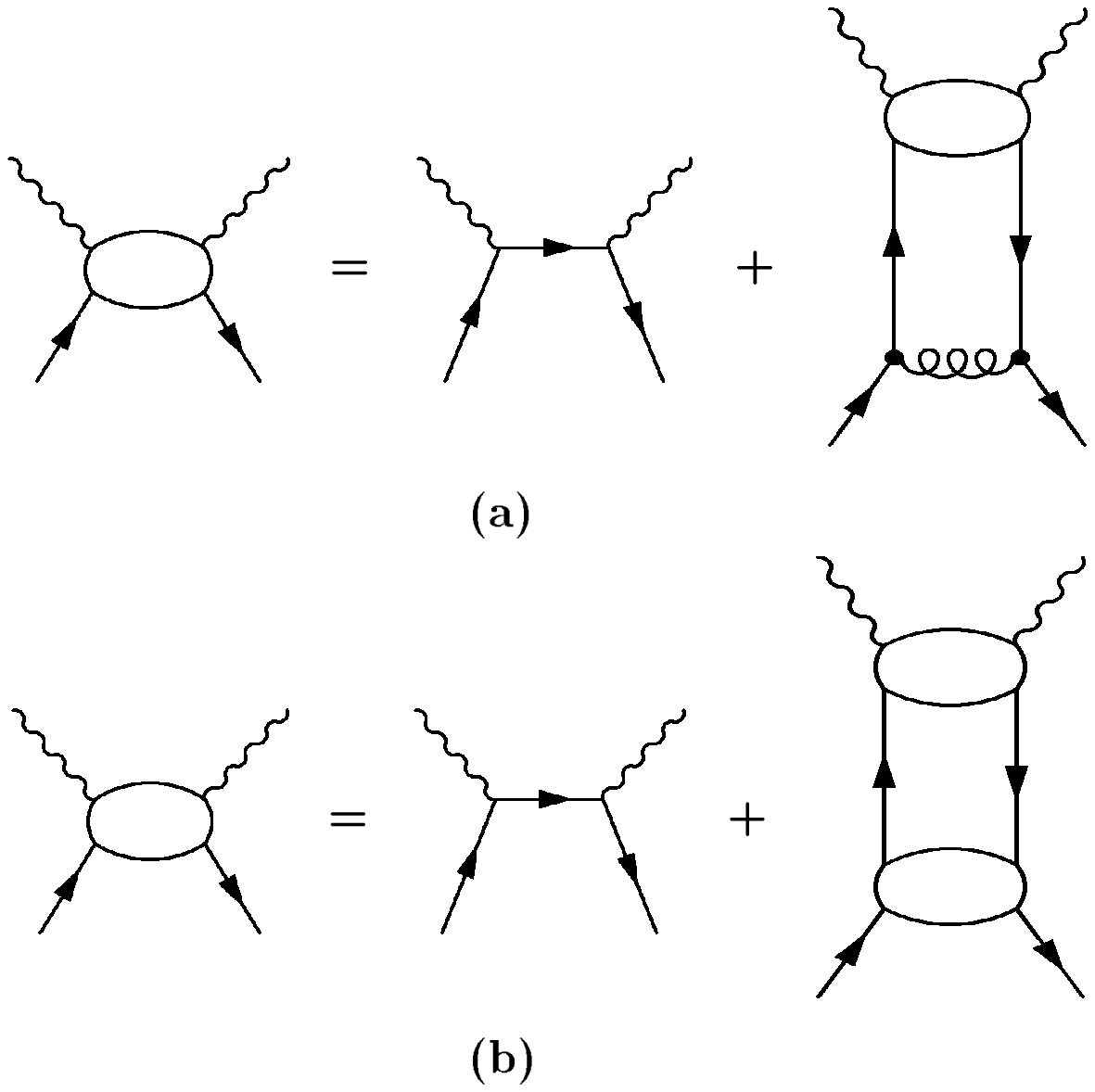}
}
\end{picture}
\caption{The evolution equation for $M$.}
\label{equation}
\end{figure}

Equation (\ref{order})
shows that only the integration over $k_{1\perp}$ has $\mu$
as the lower limit.
The lower limits for other $k_{i\perp}$ with $i \neq 1$
are expressed through the transverse momenta.
As is well known, the ordering of Eq. (\ref{order}) leads to
considering the
logarithms of $Q^2$ only. On the other hand,
as we investigate the
small-$x$ region, we will lift this ordering, allowing for the transverse
momentum of quarks in any ladder rung
to reach the lowest limit $k_{\perp} = \mu$. Obviously, this increases the
region of integration over $k_{i \perp}$.
Let us represent any graph contributing to $M$ as a convolution of
two amplitudes connected by the rung with minimal $k_{\perp}$. In the DGLAP,
such a representation is given by the graph  in Fig.\ref{ladder},
with the minimal $k_{\perp}$ in the lowest rung only; now, having lifted
the $k_{\perp}$ ordering  of Eq. $(\ref{order})$,
in the same graph, the minimal
$k_{\perp}$ is assigned to any rung. Summing up such
graphs to all loops and adding the Born graph,
one arrives at the equation represented in Fig.\ref{equation}a
if we use the DGLAP ordering and at that of Fig.\ref{equation}b
if one drops the $k_{\perp}$ ordering.
The blobs in Fig.\ref{equation} denote that radiative
corrections to all orders in QCD coupling are taken into account.
As a result of the infrared regularization we have chosen, the amplitude $M$
is $\mu$-dependent now and one can evolute $M$ with
respect to $\mu$.
It is convenient to use the Mellin transform for the amplitude $M$:

\begin{equation}
M(s/\mu^2, Q^2/ \mu^2) = \int_{-\imath\infty}^{\imath\infty}
\frac{d\omega}{2\pi\imath}
\left(\frac{s}{\mu^2}\right)^{\omega}F(\omega, Q^2/\mu^2) .
\label{mellin}
\end{equation}

As we are going to take into account both logarithms of $s/\mu^2$ and
of $Q^2/\mu^2$ in every order in $\alpha_s$, we assume that

\begin{equation}
M = M^{Born}U (\ln(s/\mu^2), \ln(Q^2/\mu^2)),
\label{U}
\end{equation}
where

\begin{equation}
M^{Born} = -\frac{s}{s - Q^2 + \imath\epsilon}
\label{Mb}
\end{equation}
and $U$ has to be properly defined.
Equation (\ref{Mb}) leads to $\Im_s M^{Born} = \pi\delta(1 - x)$, which agrees
with Eqs. (\ref{fborn}) and (\ref{defM}).

Let us note that
$U$  must have a non-zero
imaginary part, $\Im_s U$, when $s$ is positive and $\Im_s U = 0$
when $s$ is negative. This is satisfied when

\begin{equation}
U = U(\ln(- s/\mu^2), \ln(Q^2/\mu^2)) =
U((\ln(s/\mu^2) - \imath\pi), \ln(Q^2/\mu^2 )).
\label{minus}
\end{equation}

Since $\ln(s/\mu^2)$ is assumed to be much greater than $\pi$, one can
expand Eq.~(\ref{minus})
in the small-$x$ limit as follows:

\begin{equation}
M = (-1 + \imath\pi\delta(1 - x)) \left[U(\ln(s/\mu^2), \ln(Q^2/\mu^2))
-\imath\pi \frac{\partial U(\ln(s/\mu^2), \ln(Q^2/\mu^2))}
{\partial \ln(s/\mu^2)} + ...\right] ,
\label{MdM}
\end{equation}

so that

\begin{equation}
\Im_s M \approx \pi\left[\delta(1 - x)U (\ln(Q^2/\mu^2)) +
\frac{\partial U}{\partial \ln(s/\mu^2)}\right] .
\label{dM}
\end{equation}

Combining (\ref{dM}) and (\ref{mellin}),
we obtain, in the small-$x$ limit,

\begin{equation}
f_{NS} = \int_{-\imath\infty}^{\imath\infty}
\frac{d\omega}{2\pi\imath}
\left(\frac{s}{\mu^2}\right)^{\omega} \omega F_(\omega, Q^2/\mu^2) .
\label{defns}
\end{equation}

So, once we know $F$ we also know $f_{NS}$.  Now we calculate $F$ using
the equations in Fig.\ref{equation}. Differentiating the left-hand sides of the equations
in Fig.\ref{equation} with respect to $\ln\mu^2$, we obtain that

\begin{equation}
-\mu^2 \frac{\partial M}{\partial\mu^2} =
\frac{\partial M}{\partial \ln(s/\mu^2)} +
\frac{\partial M}{\partial \ln(Q^2/\mu^2)}
\label{lhs}
\end{equation}
corresponds to

\begin{equation}
\omega F + \frac{\partial F}{\partial y}
\label{lhsm el}
\end{equation}
for the Mellin amplitude $F$. We have defined $y = \ln(Q^2/\mu^2)$. \\
Then, differentiating with respect to $\ln(\mu^2)$ the right-hand
sides of the two equations in Fig.\ref{equation}, we arrive at the equation
(the Born contribution does not depend on $\mu$ and cancels under the
differentiation):

\begin{equation}
\left(\frac{\partial}{\partial y} + \omega \right)\widetilde{F}(\omega , y) =
\left[\frac{1}{8\pi^2}(1 + \lambda \omega)\right]
\widetilde{F}(\omega, y) \widetilde{F}_0^B (\omega)
\label{eqap}
\end{equation}
if we follow the DGLAP $k_{\perp}$ ordering (\ref{order}), and at the equation

\begin{equation}
\left(\frac{\partial}{\partial y} + \omega \right) F(\omega, y)  =
\left[\frac{1}{8\pi^2}(1 + \lambda \omega)\right]
F(\omega, y) F_0(\omega)
\label{eqir}
\end{equation}
if we do not invoke this $k_{\perp}$ ordering.\\

In Eqs. (\ref{eqap}) and (\ref{eqir}) we have used the
notations $\lambda = 1/2$  and provided
the DGLAP amplitudes with a tilde . The expression in the squared
brackets is the
result of a one-loop integration in the right-hand side of the
equations represented in Fig.\ref{equation}.

The Born DGLAP amplitude for the forward scattering of quarks
\begin{equation}
\widetilde{F}_0^B = 4\pi\alpha_s(Q^2)C_F/\omega,
\label{bornap}
\end{equation}
with $C_F= (N^2-1)/2N = 4/3$,
whereas the amplitude $F_0(\omega)$ for the forward scattering of quarks in
Eq. (\ref{eqir})
is unknown yet and has to be calculated independently.

Equation (\ref{eqap}) leads to the
DGLAP result for the non-singlet structure function in the leading order.
Indeed, the solution of Eq. (\ref{eqap}) is

\begin{equation}
\widetilde{F} =
\widetilde{C}
\exp\left( \int_{\mu^2}^{Q^2} \frac{dr}{r}
\frac{\alpha_s(r)C_F}{2\pi}
\left[\frac{1}{\omega} + \lambda \right] \right) ,
\label{nsap}
\end{equation}
and we therefore arrive at the famous leading-order DGLAP
formula for
$f_{NS}$:

\begin{equation}
\widetilde{f}_{NS} =  \int_{-\imath \infty}^{\imath \infty}
\frac{d\omega}{2\pi \imath} \left(\frac{1}{x} \right)^{\omega}
\widetilde{C}
\exp\left({\widetilde{A}\left[\frac{1}{\omega} + \lambda \right]}\right) ,
\label{evap}
\end{equation}
with $\widetilde{A}$ given by

\begin{equation}
\widetilde{A} = \frac{C_F}{2\pi} \int_{\mu^2}^{Q^2} \frac{d r}{r}
\alpha_s(r) .
\label{Aap}
\end{equation}

The quantity $\widetilde{C}$ is not determined a priori and has to be
fixed somehow.
However, if one
assumes the delta-function input (\ref{fborn}) corresponding to DIS off a
quark in the Born approximation, one can specify
$\widetilde{C}$  by the
matching:

\begin{equation}
\widetilde{M} = \widetilde{M^B}
\label{matchap}
\end{equation}
when $y = 0$, i.e.

\begin{equation}
\widetilde{C} = 1 .
\label{defcap}
\end{equation}

Then one can finalize the leading-order DGLAP result for the
non-singlet structure function with the delta-function input (\ref{born}) as:

\begin{equation}
\widetilde{f}_{NS} = \int_{-\imath\infty}^{\imath\infty}
\frac{d\omega}{2\pi\imath}
\exp\left(\tilde{A}{\left[\frac{1}{\omega} + \lambda \right]
}\right) .
\label{fnsap}
\end{equation}

On the other hand, in order to consider the
small-$x$ behaviour of $f_{NS}$, we have
to lift the $k_{\perp}$ ordering of Eq. (\ref{order}) and to use
Eq. (\ref{eqir}) rather than
Eq. (\ref{eqap}). The solution to (\ref{eqir}) is

\begin{equation}
F= C\exp\left({[-\omega + (1 + \lambda\omega)F_0(\omega)/8\pi^2]y}\right),
\label{cir}
\end{equation}
which contains two unknown quantities, $C$ and $F_0$, which
have to be specified.

Just as for the DGLAP, input $C$ can be fixed by making special assumptions.
Then the expression for $f_{NS}$ at small $x$ is given by
the Mellin transform of Eq. (\ref{defns}), with the integrand Eq. (\ref{cir}),
provided $F_0$ is known.
In the case of the delta-function input, one can specify $C$.
Equation (\ref{cir}) shows that $C$ actually coincides with the
Mellin amplitude
$M$ when the incoming photon
is almost on-shell, i.e. its virtuality is $\mu^2$.
So the equation for $C$ is similar to the one for $M$,
but it differs from (\ref{eqir}) on two points: first, there is no
$Q^2$-dependence in the left-hand side; second, the
Born contribution to $C$ is
$\mu$-dependent. We recall that the Born contribution to $M$ was
independent of $\mu$ (see (\ref{fborn}))  and therefore it vanished under
differentiation over $\mu$. Now we have to add the Born contribution

\begin{equation}
1/\omega
\label{born}
\end{equation}
to the right-hand side of the equation for $C$.
Thus, $C$ obeys

\begin{equation}
C(\omega) = \frac{1}{\omega} + \frac{1}{8\pi^2}[1/\omega +\lambda]
C(\omega)F_0(\omega)
\label{eqc}
\end{equation}
with the obvious solution

\begin{equation}
C= 1/[\omega - (1 + \lambda\omega)F_0(\omega)/8\pi^2]~ .
\label{defc}
\end{equation}

So, the only ingredient that still remains unknown in the right-hand
side of
Eq. (\ref{cir}) is the amplitude
$F_0$, corresponding to the  scattering of quarks.

\begin{figure}
\begin{picture}(120,180)
\put(0,10){
\epsfbox{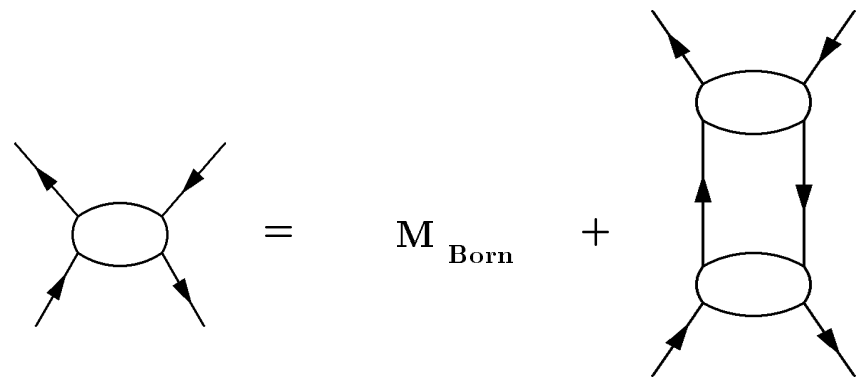}
}
\end{picture}
\caption{The evolution equation for the quark scattering amplitude.}
\label{f0}
\end{figure}

The equation for $F_0$ is shown in Fig.\ref{f0}. It looks pretty similar to
Eq. (\ref{eqc}) for $C$.
The difference is in the Born amplitude $A(\omega)/\omega$ for the
quark scattering, which
we specify below in Eq. (40).
Then, in analogy to Eq.(\ref{eqc}), we obtain

\begin{equation}
F_0(\omega) = \frac{A(\omega)}{\omega} + \frac{1}{8\pi^2}
\left[\frac{1}{\omega} +
\lambda \right] F_0^2(\omega) ,
\label{equark}
\end{equation}
which has the solution

\begin{equation}
F_0 =
4\pi^2\frac{\left[\omega -\sqrt{\omega^2 -
(1 + \lambda \omega)A(\omega)/2\pi^2} \right]}
{1 + \lambda \omega} .
\label{F0}
\end{equation}

Finally, combining the above results, we obtain the
expression for the
non-singlet structure function (cf. (\ref{nsap})):

\begin{equation}
f_{NS}= \int_{-\imath \infty}^{\imath \infty}
\frac{d \omega}{2\pi \imath}C\left(\frac{1}{x}\right)^{\omega}
\omega \exp\left([(1 + \lambda \omega)F_0/8\pi^2] y\right)
\label{evir}
\end{equation}
for an arbitrary input $C$, or

\begin{equation}
f_{NS} = \int_{-\imath\infty}^{\imath\infty} \frac{d\omega}{2\pi\imath}
\left(\frac{1}{x}\right)^{\omega}
\omega \frac{F_0(\omega)}{A(\omega)}
\exp\left([(1 + \lambda\omega)F_0/8\pi^2]y\right) ,
\label{fnsir}
\end{equation}
where we have used Eqs. (\ref{defc}) and (\ref{F0}) and $A(\omega)$
is unspecified yet.
Now let us determine $A$. In doing so we use the approach
suggested in \cite{kl}, with some important changes. \\

\begin{figure}
\begin{center}
\begin{picture}(120,150)
\put(0,10){
\epsfbox{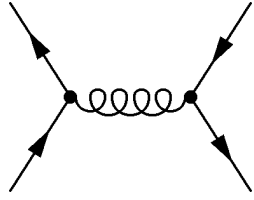}
}
\end{picture}
\end{center}
\caption{The Born graph for the quark scattering amplitude.}
\label{ferm}
\end{figure}

Let us consider the forward quark-scattering amplitude in the Born
approximation.
The only Feynman graph with a non-zero discontinuity (imaginary
part) with respect to the Mandelstam total energy
variable $s'$ of this process is
shown in
Fig.\ref{ferm}. It yields

\begin{equation}
M_0^B = -4\pi \alpha_s C_F \frac{s'}{s' + \imath \epsilon}~ ,
\label{defM0}
\end{equation}
with $\alpha_s$ fixed in the Born approximation.

\begin{figure}
\begin{picture}(120,150)
\put(0,10){
\epsfbox{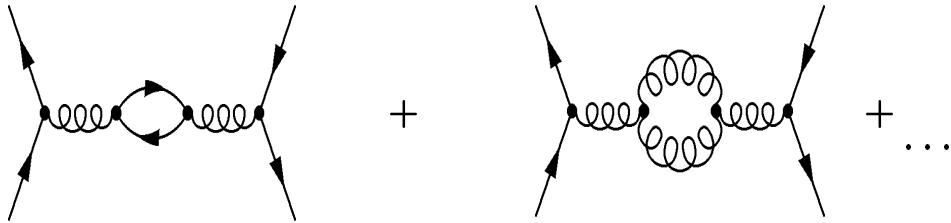}
}
\end{picture}
\caption{The graphs contributing to the running QCD coupling.}
\label{coupling}
\end{figure}

Among the radiative corrections to the Born amplitude,
there are those, see Fig.\ref{coupling}, that make
$\alpha_s$ running.
The contributions of the graphs of that kind
lead to replacing~ $M_0^B$ ~by $B$, when the QCD coupling depends
on $s'$:

\begin{equation}
B = -4\pi \alpha_s(s')C_F \frac{s'}{s' + \imath \epsilon}~ .
\label{defB}
\end{equation}

Furthermore, in the planar gauge, taking into account the radiative corrections
is equivalent to iterating $B$  (we recall that we discuss the positive
signature in the present paper). As a result of such an
iteration we arrive at the amplitude $M_0(s')$.  In order to avoid the IR
singularities in $M_0$, we again use the IR cut-off $\mu$ so
that $s' \geq \mu^2$.
in the denominators of Eqs. (\ref{defM0}) and (\ref{defB})
Equivalently, it can be expressed by
replacing $s' + \imath \epsilon$ by $s' - \mu^2 + \imath \epsilon$ in the
denominators in
(\ref{defM0}) and (\ref{defB}). One can now apply the Mellin transform to $B$:

\begin{equation}
B = \int_{-\imath \infty}^{\imath \infty} \frac{d\omega}{2\pi \imath}
\left(\frac{s}{\mu^2}\right)^{\omega} R(\omega).
\label{melR}
\end{equation}

In order to invert $R(\omega)$ through $B(s)$, one must use the inverse
transform
in Eq. (\ref{melR}).
While the Mellin transform (\ref{mellin})
is the extreme high-energy limit of the Sommerfeld~- Watson transform, the
inverse transform,  which
respects  the analytical properties of the
scattering amplitude under
discussion, involves its discontinuities
(see e.g. \cite{col}).
Then, it is appropriate to use this inverse
transform in a form similar to the one
suggested in  \cite{kl}. The inverse transform of Eq. (\ref{mellin}) is

\begin{equation}
F(\omega) = - \frac{1}{\sin(\pi \omega)}
\int_0^{\infty}d\rho \exp(-\omega \rho) \Im_s M(\rho) ,
\label{invmellin}
\end{equation}
where $\rho = \ln (s/\mu^2)$, we will approximate $\sin(\pi\omega) \approx
\pi \omega$ at small $\omega$.
Then inverting Eq. (\ref{melR}) we obtain

\begin{equation}
R(\omega)= - \frac{1}{\pi \omega} \int_0^{\infty}d\rho e^{-\omega \rho}
\Im_s B(\rho) ,
\label{defR}
\end{equation}
having defined $\rho = \ln (s/\mu^2)$.\\

Using the standard one-loop form for $\alpha_s$

\begin{equation}
\alpha_s(s) = \frac{1}{b \ln(-s/\Lambda_{QCD}^2)}~ ,
\label{alpha}
\end{equation}
when $s > 0$, we obtain

\begin{equation}
R(\omega) =  \frac{4C_F \pi}{b\omega}\int_0^{\infty} d\rho
\left[\frac{1}{(\rho + m)^2 + \pi^2} -
\frac{(\rho + m)}{[(\rho + m)^2 + \pi^2]}
\left(\frac{s}{\mu^2}\right) \delta(s/\mu^2 - 1)\right]
\exp(-\rho\omega)~ ,
\label{R}
\end{equation}
where

\begin{equation}
m = \ln(\mu^2/\Lambda_{QCD}^2)~;
\label{m}
\end{equation}
and we  assume $\mu > \Lambda_{QCD}$.
The first term in the square brackets in Eq. (\ref{R}) corresponds to the
imaginary
part of (\ref{alpha}) and the second one comes from the imaginary part of
$1/(s - \mu^2 + \imath \epsilon)$.

We rewrite the
expression for $R$ in terms of $A$ that we introduced in (\ref{equark}):

\begin{equation}
R(\omega) = \frac{A}{\omega}
\label{intA}
\end{equation}
with

\begin{equation}
A(\omega) = \frac{4C_F \pi}{b}
\left[ \frac{m}{m^2 + \pi^2} -  \int_0^{\infty}
\frac{d\rho \exp(-\rho\omega)}{(\rho + m)^2 + \pi^2}  \right]~.
\label{A}
\end{equation}

Comparing Eq. (\ref{A}) with the inverse Mellin transform of
Eq. (\ref{defM0}), we see that $A(\omega)$
corresponds to running $\alpha_s$ in the Mellin space.
In the next section we will discuss Eq. (\ref{A}) in some more detail.

\section{Asymptotics of $f_{NS}$ }

Before making use of all results obtained in the previous section, let us
dwell upon our expression for $R$ and $A$.
Equations (\ref{intA}) and (\ref{A}) differ from the similar
expressions in \cite{kl} on one essential point:
$\pi^2$ was not accounted for in either \cite{kl} or \cite{emr,m},
though its
numerical value
is quite large with respect to other terms in the integrand and cannot be
neglected a priori.

As stated above,  $A$ in (\ref{A}) stands for the QCD coupling in
the Mellin space. Indeed, if
we let $\alpha_s$ be fixed just as
in the DLA, we obtain

\begin{equation}
R_{DL} = \frac{4\pi C_F \alpha_s^{DL}}{\omega}
\label{ADL}
\end{equation}
instead of (\ref{A}). We note that in the framework of  DGLAP, we
would have obtained $\widetilde{A}$ given by (\ref{Aap}) instead of
$A/8\pi^2$ just
as it was assumed in \cite{b,jk,blum}. The reason of such a
discrepancy is obvious: in the ladder Feynman graphs, the current
argument for
$\alpha_s$ in the $n$-th ladder rung is, actually,
$\beta_{n+1}k^2_{n \perp}/\beta_n$
\footnote{$k_{n \perp}$ and $\beta$ are the standard Sudakov variables for
ladder
virtual quark momenta. For the Sudakov parametrization in the context of
$f_{NS}$, see e.g. \cite{emr,ber}.}.
Only when $x\sim 1$ do we have $\beta_{n+1} \sim \beta_n$,
and therefore  all $\beta$ can be dropped in the arguments of $\alpha_s$ in
every rung. Then, the
DGLAP ordering (\ref{order}) establishes $Q^2$ as the
upper limit of integration
over $k_{\perp}^2$. With the above approximations, one arrives in due course at
(\ref{Aap}). However, neither of those approximations holds in the
small-$x$ region, where there is no ordering in $k_{\perp}$ and the upper
limit
is $s$ rather than $Q^2$.
Dropping those approximations leads to the expression for
 $f_{NS}$ at small $x$ given by the expressions in Eqs. (\ref{evir}) and
(\ref{A}) for
an arbitrary input and by Eqs. (\ref{fnsir}) and (\ref{A})
when the delta-function input is
used.\\

It is also  easy to check that
the corresponding DGLAP expressions can be obtained
from Eqs. (\ref{evir}) and
(\ref{fnsir}). Indeed, when $x\sim1$,
the main contribution of
the Mellin factor,
$\exp [\omega \ln (1/x)]$, comes from the
region of rather large values of $\omega$, where $ \omega \ln(1/x) \leq 1$.
That makes it possible to expand the exponent $F_0$ in Eq. (\ref{evir})
into a series in $1/\omega$. Doing so, we obtain

\begin{equation}
(1 + \lambda \omega)F_0/8\pi^2 = \frac{A}{8\pi^2}
\left[\frac{1}{\omega} + \lambda \right]
+ \left(\frac{A}{8\pi^2}\right)^2 \frac{1}{\omega}
\left[ \frac{1}{\omega} + \lambda \right]^2 + O(A^3).
\label{ser}
\end{equation}

Retaining only the first term, proportional to $A$
in the right-hand side of Eq. (\ref{ser}), and
combining the result with Eq. (\ref{evir}) one arrives at
the expression for $f_{NS}$, which is similar to the DGLAP expression
in Eq. (\ref{evap}).
In order to get a complete coincidence, $y A/8\pi^2$ has to be replaced
by $\widetilde{A}$ as
given by Eq. (\ref{evap}). The same expansion of the integrand into a
$1/\omega$-series relates Eq. (\ref{fnsir}) and Eq. (\ref{fnsap}) for
$f_{NS}$ with the
delta-function input (\ref{fborn}).\\

The small-$x$ asymptotic behaviour of $f_{NS}$ can be  obtained by evaluating
$f_{NS}$ with the saddle-point method. As the asymptotics of
the integrals (\ref{evir}) and (\ref{fnsir}) are identical, we consider below
expression (\ref{evir}), which we rewrite as follows:

\begin{equation}
f_{NS} = C \int_{-\imath\infty}^{\imath\infty} \frac{d\omega}{2\pi\imath}
\exp\left[\omega \xi + \frac{y}{2}( \omega -
\sqrt{ \omega^2 -(1 + \lambda \omega)A(\omega)/2 \pi^2 }) \right] ,
\label{rewr}
\end{equation}
where we have used the notation $\xi = \ln(1/x)$ and
$y = \ln(Q^2/\mu^2)$ . According to the
saddle-point method, when $x$ tends to zero,

\begin{equation}
f_{NS} \sim f_{NS}(\omega_0) ,
\label{fas}
\end{equation}
where the stationary point $\omega_0$ is the largest of the roots
of

\begin{equation}
\frac{d}{d\omega} \left[ \omega\xi + \frac{y}{2}
(\omega - \sqrt{\omega^2 - (1 + \lambda \omega A/ 2\pi^2)}) \right] = 0 .
\label{stat}
\end{equation}

Differentiating in (\ref{stat}) leads to

\begin{equation}
[\omega^2 - (1 + \lambda \omega)A/2\pi^2]
\left(\frac{2\xi}{y} + 1 \right)^2 =
[\omega - \lambda A/4\pi^2 - (1 + \lambda\omega)(dA/d\omega)/(4\pi^2)]^2 ,
\label{eqomega}
\end{equation}
which can be solved numerically for any $\xi$, $y$. However, for
obtaining asymptotically small-$x$ behaviour a further simplification is
possible: when $\xi/y \gg 1$, the
right-hand side of Eq. (\ref{eqomega})
is small with respect to the left-hand side
and therefore $\omega_0 \approx \omega_s$,
where $\omega_s$ is the root of

\begin{equation}
\omega^2 - (1 + \lambda \omega)(A/2\pi^2) = 0.
\label{eqasomega}
\end{equation}

In other words, $\omega_0$ is the position of the leading singularity
(the branch point) of the exponent in Eq. (\ref{rewr})
in the $\omega$ plane. So, with the
square root in (\ref{rewr}) being zero when $\omega = \omega_0$, we
arrive at the power-like asymptotics:

\begin{equation}
f_{NS} \sim f_{NS}(\omega_0) =
\exp\left[ \omega_0 \xi + y \omega_0/2\right] =
\left(\frac{1}{x}\right)^{\omega_0}
\left(\frac{Q^2}{\mu^2}\right)^{\omega_0/2}  .
\label{fasy}
\end{equation}

On the other hand, if one expands the exponent in (\ref{rewr}) into a series
in $1/\omega$ (see (\ref{ser})) and accounts for an arbitrary
but  finite number of terms in the series, one arrives at the well-known DGLAP
asymptotics for $f_{NS}$ instead of the power-like one. For example,
let us retain only the first term in (\ref{ser}). In this case
$\omega_0$ is the root of

\begin{equation}
\frac{d}{d\omega}\left[ \omega\xi +
y (A/ 8\pi^2) \left(\frac{1}{\omega} + \lambda \right)\right] = 0 .
\label{statap}
\end{equation}

Replacing $(yA/8\pi^2)$ by $\tilde{A}(Q^2)$ (cf. (\ref{evap})) in order
to treat the QCD coupling as in the DGLAP,
we obtain that the leading singularity $\tilde{\omega}_0$  is now

\begin{equation}
\tilde{\omega}_0 = \sqrt{\tilde{A}/\xi}~ ;
\label{omegaap}
\end{equation}
therefore, in this approximation :

\begin{equation}
f_{NS} \sim f_{NS}(\omega_0) = e^{\sqrt{\tilde{A}\ln(1/x) }} .
\label{fasap}
\end{equation}

Thus, taking into account a finite number of NLO terms in the expressions for
the anomalous dimensions inevitably leads to the asymptotic behaviour of
the DGLAP type, whereas the power-like behaviour is the result of accounting
for the NLO contributions to all orders in $\alpha_s$.
Now let us calculate the exponent in Eq. (\ref{fasy}).
Using Eq. (\ref{A}), we eventually obtain the equation for the position of
the leading singularity $\omega_s$ when $x$ tends to zero~:

\begin{equation}
\omega^2 - (1 + \lambda \omega)\left( \frac{2C_F \pi}{b} \right)
\left[ \frac{m}{m^2 + \pi^2} - \int_0^{\infty}
\frac{d\eta e^{-\omega \eta}}{(\eta + m)^2 + \pi^2} \right] = 0.
\label{master}
\end{equation}

Here, $b$ is the first coefficient of the $\beta$-function :
$b = (33- 2n_f)/12\pi$. We recall that $m$ in (\ref{master}) stands for
$\ln (\mu^2/\Lambda_{QCD}^2)$ and that we keep
the relation $\mu > \Lambda_{QCD}$ throughout this paper.

It is easy to solve Eq. (\ref{master}) numerically, for any fixed set of
parameters
$n_f$, $\mu$, $\Lambda_{QCD}$. For example, using
$n_f = 3$, $\mu = 1~ GeV$, $\Lambda_{QCD} = 0.1~ GeV$, we obtain

\begin{equation}
\omega_s = 0.37.
\label{omegas}
\end{equation}

On the other hand, the small-$x$ behaviour of the
non-singlet structure functions could be calculated in the
double-logarithmic approximation and is given by

\begin{equation}
f_{NS} \sim x^{-\omega_{DL}}
\label{dla}
\end{equation}
with the exponent

\begin{equation}
\omega_{DL} = \sqrt{2\alpha_s^{DL} C_F/\pi}
\label{omegadl}
\end{equation}
(see \cite{emr} for details). The QCD coupling $\alpha_s^{DL}$ is of course
fixed in the DLA and its value cannot be specified within the DLA
accuracy. Nevertheless, as asymptotically the DLA is supposed to
dominate eventually over subleading
contributions, we can estimate such an ``asymptotic value''
$\alpha_s^{DL}$ by
identifying $\omega_s$ and $\omega_{DL}$. Doing so, we obtain

\begin{equation}
\alpha_s^{DL} = \omega_s^2 \pi/2C_F.
\label{fix}
\end{equation}

In particular, for the same set of parameters as given above,
we obtain $\alpha_s^{DL} = 0.16$ ~which is the value of
$\alpha_s(60 ~GeV^2)$. We note that this value is not at all related  to
the value of $\alpha_s(Q^2)$ ($Q^2$ being  the incoming photon virtuality),
which has often been used in the DLA calculations as an estimate for
the value of $\alpha_s^{DL}$.
We would like to further comment on the dependence of $\omega_s$ on the
approximation used in our calculation. First, if we had neglected the
single-logarithmic contributions, putting $\lambda = 0$ and also dropping the
$\pi^2$ terms in Eq.~(\ref{master}), we would have obtained $\omega_s = 0.43$.
It corresponds to $\alpha_s^{DL} = 0.22 = \alpha_s(6 ~Gev^2)$.
On the other hand, keeping both double- and single-logarithmic contributions
(i.e. $\lambda = 1/2$) and neglecting the $\pi^2$ terms, we would get
$\omega_s = 0.49$  and therefore, for this case,
$\alpha_s^{DL} = 0.28 = \alpha_s (1.5 ~Gev^2)$.
It is clear that the $\pi^2$ terms, which come from
analyticity, account  only for a class of constant terms and should
in principle, be neglected for consistency. However, the importance of
keeping them into account in resummation formulae has been stressed for a
long time \cite{cg}.
Unfortunately we have not been able to
compare the higher terms in the expansion of $f_{NS}$ Eq. (\ref{ser})
with fixed higher orders in $\alpha_s$ calculations since
the subleading (the single-logarithmic) contributions to $f_{NS}$,
even in the first loop
depend on the choice of the factorization and the
regularization procedures. It would be very interesting to see the effect of
SL contributions to the $f_{NS}$ with the negative signature (these
contributions appear as from the order $\alpha_s^3$).
At present, such calculations are still in progress.
Although we have called Eqs. (\ref{evir}) and (\ref{fnsir}) a
generalization of the leading-order DGLAP, those expressions actually include
also the singular  contributions in $\omega$ present in the expressions for
the two-loop DGLAP  anomalous dimension  for $f_{NS}$ as well as
other singular terms coming from the resummation of the leading-order DGLAP
anomalous dimension to all orders in $\alpha_s$.

\section{Conclusions}

In conclusion, we
have obtained a generalization of the leading-order
DGLAP for $f_{NS}$ to the case of small $x$, by resumming the leading-order
DGLAP  anomalous
dimension $1/\omega + 1/2$ to all orders in
$\alpha_s$ and by accounting for running $\alpha_s$ effects.
In addition to the double-logarithmic contributions we took into
account the single-logarithmic ones. Our result
for the first-loop contribution to $f_{NS}$ is in agreement
with \cite{cfp}, provided that identical regularizations are used.
We have demonstrated
above that our results of Eqs. (\ref{evir})   for an arbitrary input and
(\ref{fnsir}) for the delta-function input (\ref{fborn})
agree with the DGLAP when
we neglect all terms
except the first one in Eq.~(\ref{ser}).
We showed that, with
the single logarithmic contributions and  with running $\alpha_s$ taken
into account, $f_{NS}$ also has the power-like asymptotic behaviour

\begin{equation}
f_{NS} \sim x^{- \omega_s}(Q^2/\mu^2)^{\omega_s/2}
\label{fin}
\end{equation}
as was obtained earlier in the DLA \cite{emr}
at asymptotically small $x$ but with another value of the exponent.
We have also investigated the dependence
of the value of $\omega_s$ on the accuracy
of the calculations. In doing so
we obtained that subleading (single-logarithmic) terms
contribute to the asymptotic
behaviour of $f_{NS}$ by increasing
$\omega_s$ in Eq.~(\ref{fin}) from its value
$\omega_s = 0.43$ in the DLA up to $\omega_s = 0.49$.
On the contrary, the $\pi^2$-terms strongly suppress that growth,
decreasing the value of $\omega_s$
down to $\omega_s = 0.37$. This result can be regarded as an indication for
the importance of the subleading contributions.
In the present work we have discussed mainly the asymptotic behaviour
of $f_{NS}$. A detailed comparison of our results with the DGLAP at small
but finite values of $x$ will be given elsewhere.

\section{Acknowledgements}
We are grateful to G. Altarelli, S. Catani and S. Forte for useful discussions.
This work suppoorted in part by the EU QCDNET contract FMRX-CT98-0194.

\centerline{\bf Figure captions}

Fig.1~~The Born graphs for the amplitude $M$.\\

Fig.2~~The graphs contributing to $f_{NS}$ in one-loop approximations.\\

Fig.3~~A typical graph for $f_{NS}$ in the planar gauge.\\

Fig.4~~The evolution equation for $M$.\\

Fig.5~~The evolution equation for the quark scattering amplitude.\\

Fig.6~~The Born graph for the quark scattering amplitude.\\

Fig.7~~The graphs contributing to the running QCD coupling.\\

\newpage


\end{document}